\documentclass[twocolumn,showpacs,preprintnumbers,amsmath,amssymb]{revtex4}

\usepackage{graphicx}
\usepackage{dcolumn}
\usepackage{bm}

\begin{document}

\title{Bounds to unitary evolution}

\author{Mark Andrews}
\email{Mark.Andrews@anu.edu.au}
\affiliation{Department of Physics, Faculty of Science, Australian National University, ACT 0200, Australia.}

\date{\today}

\begin{abstract}
Upper and lower bounds are established for the survival probability $|\langle\psi(0)|\psi(t)\rangle|^{2}$ of a quantum state, in terms of the energy moments $\langle\psi(0)|H^{n}|\psi(0)\rangle$. Introducing a cut-off in the energy generally enables considerable improvement in these bounds and allows the method to be used where the exact energy moments do not exist.
\end{abstract}

\pacs{03.65.-w}

\maketitle

\section{\label{sec:intro}Introduction}

How rapidly (or how slowly) can a state evolve? This question has been approached \cite{MT,FGR,Bhatt,Pf2} through the survival probability of the state: 
\begin{equation}
\label{eq:Pdef}
P(t):=|\langle \psi |\exp(-\imath H t/\hbar)\psi\rangle|^{2}=|\langle \exp(-\imath H t/\hbar)\rangle|^{2},
\end{equation}
which is the probability that the system, initially in state $\psi$, will be found to be still in that state after time $t$. The Hamiltonian $H$ is assumed to be independent of the time. Lower bounds to $P(t)$ have been established \cite{MT,Bhatt} in terms of the energy uncertainty $\Delta E:=\langle (H-\langle H\rangle)^{2}\rangle ^{1/2}$, but these bounds can be well below the actual evolution. Also it has been claimed \cite{Pf1} that there can be no upper bound (other than $P(t)\leqslant 1$) in terms of $\Delta E$ alone; without an upper bound one can not be sure that the state will change at all. Here, upper and lower bounds will be found in terms of higher energy moments. 

Expanding the exponential in Eq.(\ref{eq:Pdef}) as a power series gives
\begin{equation}
\label{eq:Pseries}
P(t)=1-(h_{2}-h_{1}^{2})\frac{t^{2}}{\hbar^{2}}+(h_{4}-4h_{3}h_{1}+3h_{2}^{2})\frac{t^{4}}{12\hbar^{4}}-...,
\end{equation}
where $h_{n}:=\langle H^{n}\rangle$, the $n$-th energy moment. The absence of a linear term in this expansion is important to the discussion of the quantum Zeno effect \cite{Peres,Merz}.

There are good reasons why $h_{1}$ and $h_{2}$ should exist \cite{FGR}, but for many states used in physics some of the higher energy moments do not exist. It will be established in Section {\ref{sec:bounds}} that, apart from the factor $(-1)^{n}$, the coefficient of $t^{2n}$ in Eq.(\ref{eq:Pseries}) must be positive if it exists, and that the partial sums of this series give alternately upper and lower bounds to $P(t)$; an upper bound if the last included term is positive or a lower bound if the last included term is negative. Section {\ref{sec:cutoff}} introduces a cut-off in the energy, equivalent to projecting onto a finite-energy subspace. This enables considerable improvement in these bounds (including the one in terms of $\Delta E$) and also allows the method to be used where the exact energy moments do not exist. We first need to show that the coefficients in Eq.(\ref{eq:Pseries}) are moments over the autocorrelation of the energy distribution.  

\section{\label{sec:EnergyDistrn}The energy distribution}

Consider a complete set of commuting observables $H, K$ with common eigenstates $|\phi_{E,\kappa}\rangle$, so that $H |\phi_{E,\kappa}\rangle=E |\phi_{E,\kappa}\rangle$ and $K |\phi_{E,\kappa}\rangle=\kappa |\phi_{E,\kappa}\rangle$. (In general, $K$ represents a set of operators and $\kappa$ a set of eigenvalues.)  Then, for any function $f(H)$ of the Hamiltonian,
\begin{equation}
\label{eq:f}
\langle f(H) \rangle = \int\!\! dE \! \int \!\! d\kappa\,f(E)|\langle \phi_{E,\kappa}|\psi \rangle |^{2}.
\end{equation}
Define the energy distribution $\rho(E):=\int d\kappa |\langle \phi_{E,\kappa}|\psi \rangle |^{2}$. Then $\int dE\,\rho(E)=1$ and
\begin{equation}
\label{ }
\langle f(H) \rangle = \int dE\,\rho(E)\,f(E).
\end{equation}
For example, $h_{n}:=\langle H^{n}\rangle = \int dE\,\rho(E)\,E^{n}$.

Let $L$ and $M$ be the lower and upper bounds to the energies for which $\rho(E)$ is non-zero, but allow the possibility that $M=\infty$ and even the unphysical case that $L=-\infty$. The survival probability $P(t)$ is not changed by a shift in energy and for all the examples used here $L=0$ or $-\infty$. In terms of $\rho(E)$,
\begin{eqnarray}
\label{ }
P(t) & = & \int_{L}^{M} \!\!dE\int_{L}^{M} \!\! dE' \rho(E)\rho(E') e^{\imath(E'-E)t/\hbar}   \nonumber \\ 
 & = & \int_{L}^{M} \!\! dE\int_{L}^{M} \!\! dE' \rho(E)\rho(E') \cos\left(\frac{E'-E}{\hbar}t\right)\!\!. 
\end{eqnarray}
Change the integration variables from $\{E, E'\}$ to $\{E, \epsilon\}$ with $\epsilon :=E'-E$, and introduce the autocorrelation of $\rho(E)$ through the even function
\begin{equation}
\label{eq:W}
W(\epsilon):=2\int_{L}^{M-\epsilon} dE\, \rho(E)\rho(E+\epsilon)\;\; \text{for}\; \epsilon > 0.
\end{equation}
Then $W(\epsilon)$ is never negative and
\begin{equation}
\label{eq:P-W}
P(t)=\int_{0}^{M-L} d\epsilon \,W(\epsilon)\cos(\frac{\epsilon\, t}{\hbar}).
\end{equation}
Also $\int_{0}^{M-L} d\epsilon \,W(\epsilon)=P(0)=1$.

Expanding the cosine in Eq.(\ref{eq:P-W}), gives
\begin{equation}
\label{eq:P-e}
P(t)= 1-\frac{e_{2}t^{2}}{2!\hbar^{2}}+ \frac{e_{4}t^{4}}{4!\hbar^{4}}- \frac{e_{6}t^{6}}{6!\hbar^{6}}+...,
\end{equation}
[which must agree with Eq.(\ref{eq:Pseries})], where
\begin{equation}
\label{eq:en}
e_{n}:=\int_{0}^{M-L} d\epsilon \,W(\epsilon)\epsilon^{n}.
\end{equation}

The moments $e_{n}$ are positive and can be expressed, for even $n$, in terms of the energy moments $h_{k}$ with $k \leq n$\,:
\begin{eqnarray}
e_{n} & = & \frac{1}{2}\int_{L-M}^{M-L}\!\!d\epsilon \,W(\epsilon)\epsilon^{n}=\int\!\!\!\int \!\!d\epsilon \,dE \, \rho(E)\rho(E+\epsilon)\epsilon^{n} \nonumber \\
 & = & \int\!\!\!\int \!\!dE\, dE' \, \rho(E)\rho(E')(E-E')^{n}. \label{eq:e-E}
\end{eqnarray} 
Now expand $(E-E')^{n}$ as a sum of products of powers of $E$ and $E'$. Thus, for example, $(E-E')^{2}=E^{2}-2EE'+E'^{2}$ leads to $e_{2}=h_{2}-2h_{1}^{2}+h_{2}=2(h_{2}-h_{1}^{2})=2(\Delta E)^{2}$ and $e_{4}=2(h_{4}-4h_{3}h_{1})+6h_{2}^{2}$. More generally, 
\begin{equation}
\label{eq:enh}
\frac{e_{n}}{n!}=2\sum_{k=0}^{n/2-1}(-1)^{k}\frac{h_{k}}{k!}\frac{h_{n-k}}{(n-k)!}+(-1)^{n/2}(\frac{h_{n/2}}{(n/2)!})^{2}.
\end{equation}

\section{\label{sec:bounds}Bounds on the survival}

We first show that the partial sums of the Taylor series for $\cos x$ provide alternately upper and lower bounds to $\cos x$. To prove this, note that if we have an upper (or lower) bound to $\cos x$ for all $x\geq 0$, applying $\sin x=\int_{0}^{x}\cos u \,du$ gives an upper (lower) bound to $\sin x$. Then applying $\cos x=1-\int_{0}^{x}\sin u \,du$ gives a lower (upper) bound to $\cos x$. Thus $\cos x  \leq 1\Rightarrow \sin x \leq x \Rightarrow \cos x  \geq 1-\frac{1}{2}x^{2} \Rightarrow \sin x \geq x -\frac{1}{3!}x^{3}\Rightarrow \cos x  \leq 1-\frac{1}{2!}x^{2}+ \frac{1}{4!}x^{4}$ and so on. Using this sequence of inequalities for $\cos x$, it follows directly from Eq.(\ref{eq:P-W}) that 
\begin{eqnarray}
P(t) & \geq & 1-\frac{e_{2}t^{2}}{2\hbar^{2}}=1-(\Delta E)^{2}\frac{t^{2}}{\hbar^{2}} \label{eq:P2}\\
P(t) & \leq & 1-\frac{e_{2}t^{2}}{2!\hbar^{2}}+ \frac{e_{4}t^{4}}{4!\hbar^{4}}\label{eq:P4}\\
P(t) & \geq & 1-\frac{e_{2}t^{2}}{2!\hbar^{2}}+ \frac{e_{4}t^{4}}{4!\hbar^{4}}- \frac{e_{6}t^{6}}{6!\hbar^{6}}\label{eq:P6}
\end{eqnarray}
and so on. Whereas the series Eq.(\ref{eq:P-e}) for $P(t)$ may or may not converge, each of these bounds is valid provided the moments $e_{n}$ in it exist. 

The lower bound in Eq.(\ref{eq:P2}) is well known and has been improved \cite{MT,Bhatt} to $P(t) \geq \cos^{2}(\Delta E\,t/\hbar)$. To my knowledge, the other bounds are new; furthermore I know of no other upper bounds to $P(t)$.

\begin{figure}
\includegraphics[]{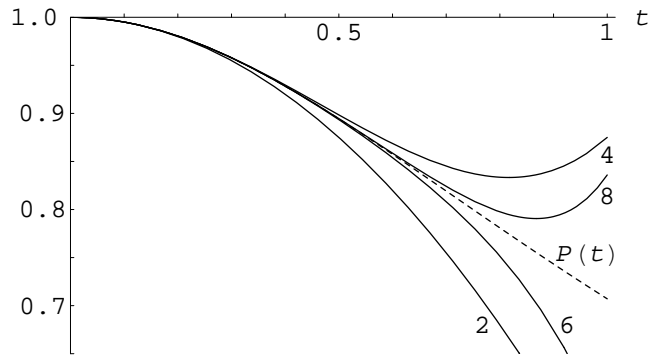}
\caption{\label{fig:gauss} Bounds to the survival probability for the energy distribution $\rho (E) \propto E^{-1/2}e^{-E /\gamma},\,E>0$. The time unit is $\hbar /\gamma $. The dashed curve is the exact survival probability $P(t)$ and the solid curves are the bounds given by Eqs.(\ref{eq:P2}-\ref{eq:P6}). The numbers on these bounds give the largest energy moment used.}
\end{figure}
As a simple example, consider the energy distribution $\rho (E) \propto E^{-1/2}e^{-E/\gamma}$, $E>0$. [One of many possible realizations of this as a wavefunction is as the free Gaussian $\psi = (t-\imath \tau)^{-1/2}\,\exp (\frac{1}{2\hbar}\imath m x^{2}/(t-\imath \tau))$, where $\tau =\frac{1}{2}\hbar/\gamma$.] The exact evolution has the survival probability $P(t)=(1+(\gamma t/\hbar)^{2})^{-1/2}$ and the coefficient of $t^{n}$ in the power series for this must be $e_{n}/n!$. In fact, $e_{n}=(n-1)^{2}(\gamma /\hbar)^{2}e_{n-2}$ with $e_{0}=1$. The autocorrelation is $W(\epsilon )=2/(\pi \gamma)\,K_{0}(\epsilon /\gamma)$, but it is not needed. All energy moments exist because the energy distribution falls off faster than any power of the energy. Fig.~\ref{fig:gauss} shows the exact evolution and four successive bounds for this case.

\begin{figure}
\includegraphics[]{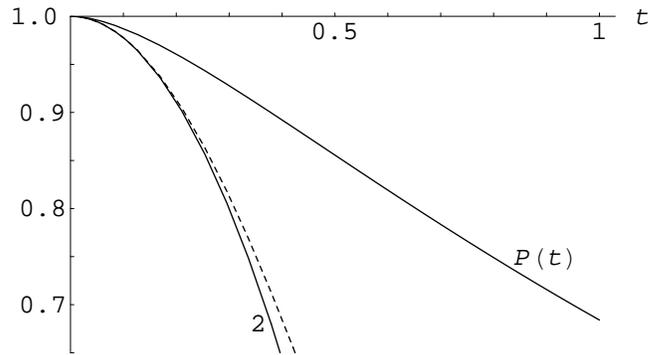}
\caption{\label{fig:7-2} Bounds to the survival probability for the energy distribution $\rho (E)\propto\ (1+E/\gamma)^{-7/2},\,E>0$. The upper curve is the exact survival probability $P(t)$ and the solid curve labelled ``2'' shows the bound given by Eq.(\ref{eq:P2}). The dashed curve is $\cos^{2}(\Delta E\, t/\hbar)$, which is the best possible lower bound if only $\Delta E$ is known. The time unit is $\hbar /\gamma$.}
\end{figure}
When the energy distribution $\rho (E)$ decreases slowly (slower than exponentially) then higher energy moments may not exist and even if some do exist the bounds provided by Eqs.(\ref{eq:P2}-\ref{eq:P6}) may be very poor. For example, if $\rho (E)\propto\ (1+E/\gamma)^{-7/2}$, $E>0$, then $e_{2}=\frac{40}{9}\gamma^{2}$ but no other even moments exist. Eqs.(\ref{eq:P2}-\ref{eq:P6}) provide no upper bound and the lower bound $P(t)\leq 1-\frac{1}{2}e_{2}t^{2}/\hbar^{2}$ is very poor, as shown in Fig.~\ref{fig:7-2}. We will now show how using an energy cut-off yields good upper and lower bounds for this system and many others.

\section{\label{sec:cutoff}Using a cut-off in the energy}

Following the work of Uffink and Hilgevoord \cite{U}, we cut off the energy at say $E=c$ and write
\begin{equation}
\label{ }
\alpha :=\int_{L}^{c}\rho (E)\,dE. 
\end{equation}
The exact state of the system can be expressed as 
\begin{equation}
\label{ }
\psi (t)=\sqrt{\alpha}\,\bar{\psi}(t)+\sqrt{1-\alpha}\,\chi(t),
\end{equation}
where, in the notation of Eq.(\ref{eq:f}),
\begin{eqnarray}
\!\bar{\psi}(t)\!\! & := &\!\! \frac{1}{\sqrt{\alpha}}\int d\kappa \int_{L}^{c} dE\, e^{-\imath Et/\hbar}\rho (E)\phi_{E,\kappa}\,, \\
\!\chi(t)\!\! & := &\!\! \frac{1}{\sqrt{1-\alpha}}\int d\kappa \int_{c}^{M} dE\, e^{-\imath Et/\hbar}\rho (E)\phi_{E,\kappa}. 
\end{eqnarray}
Then $\bar{\psi}(t)$ and  $\chi(t)$ are normalized and orthogonal, and the survival amplitude $A(t):=\langle \psi (0) |\psi (t)\rangle$ is
\begin{equation}
\label{ }
A(t)=\alpha \bar{A}(t) + (1-\alpha)B(t),
\end{equation}
where $\bar{A}(t):=\langle \bar{\psi} (0) |\bar{\psi} (t)\rangle$ and $B(t):=\langle \chi (0) |\chi (t)\rangle$. Since $|B(t)|\leq 1$ we have the inequalities \cite{U}
\begin{eqnarray}
|A(t)| & \leq & \alpha |\bar{A}(t)| + (1-\alpha) \\
|A(t)| & \geq & \alpha |\bar{A}(t)| - (1-\alpha).
\end{eqnarray}
 We now apply the bounds in Eqs.(10)-(12) to $|\bar{A}|$:
 \begin{eqnarray}
\!\!|A(t)|\! & \geq & \!\alpha (1-\frac{\bar{e}_{2}t^{2}}{2\hbar^{2}})^{1/2}-(1-\alpha)\label{eq:A2} \\
\!\!|A(t)|\! & \leq & \! \alpha (1-\frac{\bar{e}_{2}t^{2}}{2!\hbar^{2}}+ \frac{\bar{e}_{4}t^{4}}{4!\hbar^{4}})^{1/2}+ (1-\alpha)\label{eq:A4}\\
\!\!|A(t)|\! & \geq & \!\alpha (1-\frac{\bar{e}_{2}t^{2}}{2!\hbar^{2}}+ \frac{\bar{e}_{4}t^{4}}{4!\hbar^{4}}- \frac{\bar{e}_{6}t^{6}}{6!\hbar^{6}})^{1/2}\! - (1-\alpha)\label{eq:A6}
\end{eqnarray}
and so on, where\begin{eqnarray}
\bar{e}_{n} & := & \int_{0}^{c-L} \!\!d\epsilon \,\overline{W}(\epsilon)\epsilon^{n},\label{eq:ebar} \\
\overline{W}(\epsilon) & := & \frac{2}{\alpha^{2}}\int_{L}^{c-\epsilon} \!\! dE\,\rho(E)\rho(E+\epsilon). 
\end{eqnarray}
Note that it is not necessary to calculate the autocorrelation $\overline{W}$ for the truncated system; the moments $\bar{e}_{n}$ can be obtained from the energy moments of the truncated system using Eq.(\ref{eq:enh}).

\begin{figure}
\includegraphics[]{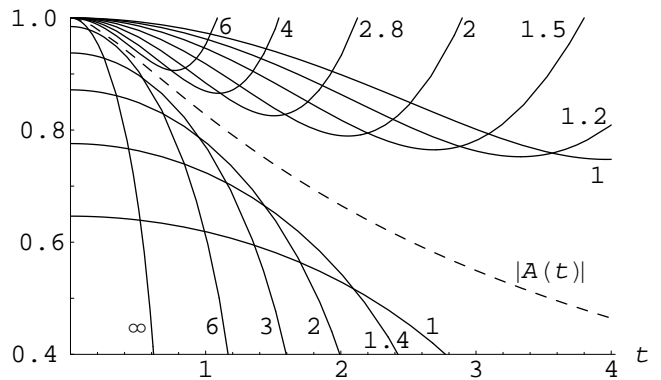}
\caption{\label{fig:7-2seq} Bounds to the survival magnitude for the energy distribution $\rho (E)\propto\ (1+E/\gamma)^{-7/2},\,E>0$. The dashed curve is the exact magnitude $|A(t)|$ of the survival and the solid curves are examples of the bounds in Eqs.(\ref{eq:A2}-\ref{eq:A4}) for selected values of the cut-off energy. The lower bounds come from the quadratic form in Eq.(\ref{eq:A2}) and the upper from the quartic form in Eq.(\ref{eq:A4}).  The numbers on these curves give the values of $c$ in units of $\gamma$. The curve with $c=\infty$, i.e. no cut-off, corresponds to the quadratic form in Eq.(\ref{eq:P2}); this also appears in Fig.~\ref{fig:7-2}}.
\end{figure}
Returning to the example with $\rho (E)\propto\ (1+E/\gamma)^{-7/2}$ which has $\alpha=1-(1+c/\gamma)^{-5/2}$, Fig.~\ref{fig:7-2seq} shows these bounds for selected values of $c$ using the quadratic lower bound in Eq.(23) and the quartic upper bound in Eq.(24).

\begin{figure}
\includegraphics[]{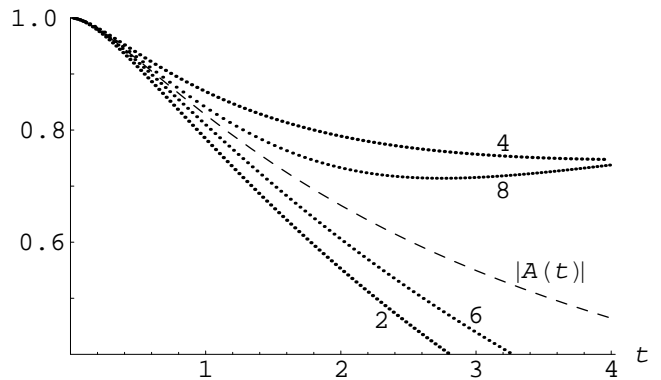}
\caption{\label{fig:7-2env} Bounds for the same system as in Fig.~\ref{fig:7-2seq}. The dotted curves are the envelopes of the bounds given by Eqs.(\ref{eq:A2}-\ref{eq:A6}). The numbers on these envelopes give the largest energy moment used.}
\end{figure}

It is now clear that the bounds from a given cut-off $c$ are good only for a limited range of the time and the best we can do with each of the bounds in Eqs.(\ref{eq:A2}-\ref{eq:A6}) is to calculate the envelope as $c$ varies. Each bound to $|A(t)|$ is given by a function
\begin{equation}
\label{eq:y}
y(t,c):=\alpha (c) \sqrt{p_{n}(t,c)}\pm (1-\alpha (c)),
\end{equation}
where $p_{n}(t,c)$ is a polynomial of degree $n$ in $t$. The envelope of these bounds as $c$ varies is found by solving
\begin{equation}
\label{eq:env}
\partial_{c}\,y(t,c)=0
\end{equation}
to give $t$ as a function of $c$. This can be done explicitly for the quadratic case, $n=2$, and for the quartic, $n=4$; but in general only numerical solution is practical. The details are in the appendices. Then inserting this $t(c)$ into $y(t,c)$ gives the envelope $y(t(c),c)$ at time $t(c)$ parametrically in terms of $c$. The envelopes for the first four bounds for the distribution $\rho (E)\propto\ (1+E/\gamma)^{-7/2}, E>0$ are shown in Fig.~\ref{fig:7-2env}.

\begin{figure}
\includegraphics[]{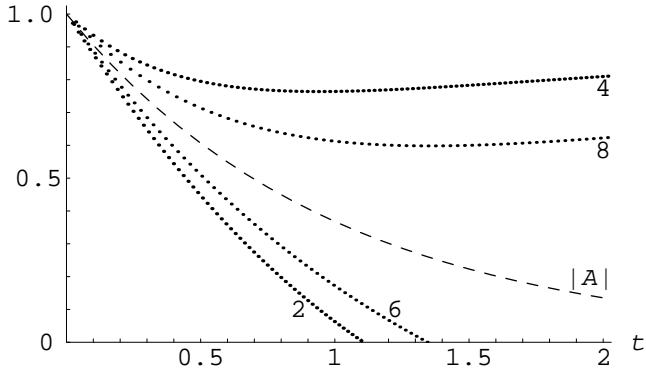}
\caption{\label{fig:BW} Bounds for the Breit-Wigner distribution. The solid curve is the exact magnitude $|A(t)|=e^{-\gamma t/\hbar}$ of the survival, and the dotted curves are the envelopes of the bounds given by Eqs.(\ref{eq:A2}-\ref{eq:A6}). The numbers on these bounding curves give the largest energy moment used. The time units are $\hbar /\gamma$.}
\end{figure}

\pagebreak  %
The Breit-Wigner system $\rho (E)\propto[(E-E_{0})^{2}+\gamma^{2}]^{-1}$ is unphysical because neither $\langle H\rangle$ nor $\langle H^{2}\rangle$ exist, but it is used because it exhibits exact exponential decay: $|A(t)|=e^{-\gamma t/\hbar}$. None of the bounds in Eqs.(\ref{eq:P2}-\ref{eq:P6}) can be used, but cutting off all energies outside the range $E_{0}-c$ to $E_{0}+c$ gives the bounds shown in Fig.~\ref{fig:BW}.

\section{\label{sec:real-imag}Bounds on the real and imaginary parts of the survival amplitude}

\begin{figure}
\includegraphics[]{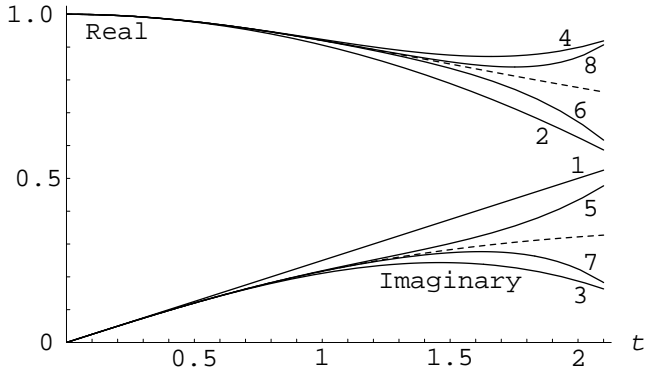}
\caption{\label{fig:RI} Bounds for the real part and the negative of the imaginary part of the survival amplitude $A(t)$ for the distribution $\rho (E) \propto E^{-1/2}e^{-E /\gamma},\,E>0$. The dashed curves show the exact amplitude and the solid curves are the bounds given by Eqs.(\ref{eq:I1}-\ref{eq:R4}). The numbers on these curves give the largest energy moment used.}
\end{figure}

The survival amplitude can be expressed as $A(t)=\int dE \rho (E) \exp (-\imath E\,t/\hbar)=R(t)-\imath \,I(t)$. The methods used in Section {\ref{sec:bounds}} applied to $R(t)$ and $I(t)$ lead to
\begin{eqnarray}
I(t) & \leq & \frac{h_{1}t}{\hbar} \label{eq:I1}\\
R(t) & \geq & 1-\frac{h_{2}t^{2}}{2\hbar^{2}} \label{eq:R2}\\
I(t) & \geq & \frac{h_{1}t}{\hbar}- \frac{h_{3}t^{3}}{3!\hbar^{3}}\label{eq:I3}\\
R(t) & \leq & 1-\frac{h_{2}t^{2}}{2!\hbar^{2}}+ \frac{h_{4}t^{4}}{4!\hbar^{4}}\label{eq:R4},
\end{eqnarray}
and so on. Fig.~\ref{fig:RI} shows the results when this is applied to the same distribution as used in Fig.~\ref{fig:gauss}. Again an energy cut-off could be used to improve these bounds, or to apply them when the energy moments do not exist.

\section{\label{sec:discuss}Discussion}

All cases considered here have just one $t(c)>0$ from Eq.(\ref{eq:appenv}) for each $n$, but this has not been proven in general. Note that if $\rho(E)$ increases suddenly then $t(c)$ may not be monotonic in $c$; then there is more than one bound for a period of time and the only the best one is of value.  

Thanks are due to M J W Hall for useful comments.

\appendix
\section{Determining the envelope}

Applying $\partial_{c}\,y(t,c)=0$ to $y(t,c)$ in Eq.(\ref{eq:y}) leads to $p_{n}(t,c)=(p_{n}(t,c)+\frac{\alpha}{2\alpha'}\partial_{c}\,p_{n}(t,c))^{2}$. From Eq.(\ref{eq:ebar}), $\partial_{c}\,\bar{e}_{n}=-2 (\alpha' / \alpha)(\bar{e}_{n}-b_{n})$, where $\alpha':=\partial_{c}\alpha = \rho (c)$,
\begin{equation}
\label{ }
b_{n}:=\frac{1}{\alpha}\int_{L}^{c}\!\!\!dE\,\rho (E)(c-E)^{n}=\sum_{k=0}^{n}\frac{(-1)^{k}n!}{k!(n-k)!}c^{k}\bar{h}_{n-k}
\end{equation}
and $\bar{h}_{k}:=\alpha^{-1}\int_{L}^{c}dE\,\rho (E)E^{n}$. Then the equation for the envelope is
\begin{equation}
\label{eq:appenv}
\sum_{\text{even}\,k}^{n}(-1)^{k/2}\frac{\bar{e}_{k}t^{k}}{\hbar^{k}k!}=\big{(}\sum_{\text{even}\,k}^{n}(-1)^{k/2}\frac{b_{k}t^{k}}{\hbar^{k}k!}\;\big{)}^{2}.
\end{equation} 
This must be solved for $t(c)>0$ and the envelope is $p_{n}(t(c),c)$ at time $t(c)$.

Replacing $E-E'$ by $(c-E')-(c-E)$ in Eq.(\ref{eq:e-E}) shows that $\bar{e}_{n}/n!=\sum_{k=0}^{n}(-1)^{k}B_{k}B_{n-k}$, where $B_{k}:=b_{k}/k!$; so the envelope equation is simplest in terms of the $B_{k}$ only. For the \textbf{quadratic} bound ($n=2$): $t(c)=2\hbar b_{1}/b_{2}$. The \textbf{quartic} bound ($n=4$) leads to a cubic equation in $t^{2}$: $B_{1}^{2}-2B_{1}B_{3}\eta+2B_{2}B_{4}\eta^{2} -B_{4}^{2}\eta^{3}=0$ with $\eta:=t^{2}/\hbar^{2}$. The relevant solution of this is $t(c)^{2} = 8\hbar^{2}( b_{2} - d_{2} + d_{1}/d_{2})/b_{4}$, where $d_{1}:=b_{1}b_{3}-b_{2}^{2}$, $d_{2}:=((d_{1}^{3}+d_{3}^{2})^{1/2}-d_{3})^{1/3}$, with $d_{3}:=(16 b_{2}^{3}-24 b_{1}b_{2}b_{3}+9 b_{1}^{2}b_{4})/16$. Note that all the $b_{k}>0$, and $d_{1}>0$ from the Schwarz inequality. Also $16 d_{3}=(4b_{2}^{3/2}-3b_{1}b_{4}^{1/2})^{2}+24b_{1}b_{2}(b_{2}^{1/2}b_{4}^{1/2}-b_{3})$ and this is positive because $b_{2}b_{4}>b_{3}^{2}$ from the Schwarz inequality. It follows that this root of the cubic is positive. The next bound $(n=6)$ leads to a quintic equation in $t^{2}$ and numerical solution is probably the most practical option.

\section{finite range of energies}

When $\rho (E)=0$ for all $E>M$, the energy cut-off must also stop at $c=M$. This causes the envelope to be valid only for times greater than $t(M)$ given by $\partial_{c}y=0$ at $c=M$. But the envelope will match smoothly to the bound without cut-off at that time, because the envelope osculates the sequence of bounds as $c$ approaches $M$. These two bounds together provide a continuous bound for all times until the envelope reaches either $0$ or $1$.

To illustrate this, consider the simple case of a square energy distribution: $\rho (E) =1/M$ for $0\leqslant E \leqslant M$. With an energy cut-off at $E=c$, $\bar{e}_{n}/n!=2c^{n}/(n+2)!$ and $b_{n}/n!=c^{n}/(n+1)!$, so the envelope equation (\ref{eq:appenv}) becomes independent of $c$ if $t(c)=\hbar \tau_{n}/c$, where $\tau_{n}$ is a dimensionless constant. Each upper envelope has the form $1-(1-\sigma_{n})t_{n}/t$ and each lower envelope the form $(1+\sigma_{n})t_{n}/t\,-1$, where $\tau_{n}$ is the positive solution of
\begin{equation}
\label{eq:tau}
\sum_{\text{even}\,k}^{n}\frac{(-1)^{k/2}2\tau_{n}^{k}}{(k+2)!}=\big{(}\sum_{\text{even}\,k}^{n}\frac{(-1)^{k/2}\tau_{n}^{k}}{(k+1)!}\;\big{)}^{2},
\end{equation}
$t_{n}:=\hbar \tau_{n}/M$ and $\sigma_{n}^{2}$ equals either side of Eq.(\ref{eq:tau}).

\begin{figure}
\includegraphics[]{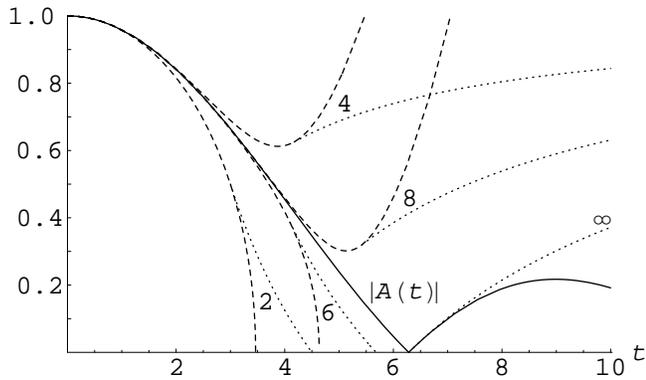}
\caption{\label{fig:sq} Bounds for the survival magnitude $|A(t)|$ for $\rho (E) =1/M,\,0<E\leqslant M$. The solid curve show the exact amplitude, the dashed curves are the bounds for $n=2, 4, 6, 8$ and the dotted curves are the corresponding envelopes for varying cut-offs. The dotted curve labeled $\infty$ is the limit of the upper envelopes as $n\rightarrow \infty$. The unit of time is $\hbar/M$.}
\end{figure}

The envelope for $n=2$ has $\tau_{2}=3$ and  $\sigma_{2}=1/2$, which gives the bound to $|A(t)|$ as $y_{2}=9\hbar/(2Mt)-1$ for $t>3\hbar/M$. Since $y_{2}=0$ at $Mt/\hbar=9/2$, the useful range of this envelope is $3<Mt/\hbar <4\frac{1}{2}$.  This can be seen in Fig.~\ref{fig:sq}, which also shows the bounds for $n=4, 6, 8$. The limit as $n\rightarrow \infty$ can be taken for the upper bounds by summing the series in Eq.(\ref{eq:tau}) in terms of $\sin \tau_{n}$ and $\cos \tau_{n}$, leading to $\tau_{\infty}=2\pi$, $\sigma_{\infty}=0$ and the bound $|A(t)|< 1-2\pi\hbar/(Mt)$ for $t>2\pi\hbar/M$.

\section{discrete energy spectra}

The envelope does not exist in regions where the energy spectrum is discrete, i.e. where the energy distribution consists of $\delta$-functions only, because the bounds provided by Eqs.(\ref{eq:A2})-(\ref{eq:A6}) do not change as $c$ moves from one $\delta$-function to the next. Then the present method gives a continuous series of bounds and $t(c)$ can be used to specify the period of time that each bound will be valid. 

\begin{figure}
\includegraphics[]{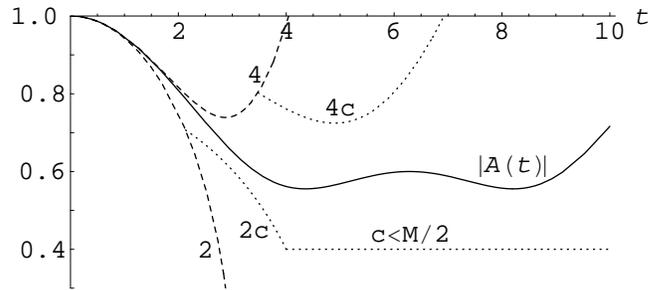}
\caption{\label{fig:3deltas} Bounds for the survival magnitude $|A(t)|$ for $\rho (E) =0.7\,\delta (E) +0.2\,\delta (E-M/2)+0.1\,\delta (E-M)$. The solid curve shows the exact amplitude, the dashed curves labeled 2 and 4 are the bounds without cut-off, the dotted curves labeled 2c and 4c are the bounds with $M/2<c<M$, and the line for $c<M/2$ is also shown. The unit of time is $\hbar/M$.}
\end{figure}

As a simple example, consider a 3-state system with equal space between the energy levels: $\rho (E)=a_{0}\delta (E) +a_{1}\delta (E-M/2)+a_{2}\delta (E-M)$ with $a_{0}+a_{1}+a_{2}=1$. For $0<c<M/2$, $\alpha = a_{0}$, all the $h_{k}$ are zero, and the only bound that can be obtained from Eqs.(\ref{eq:A2})-(\ref{eq:A6}) is $|A(t)|\geqslant 1-2a_{0}$ (useful only if $a_{0}<\frac{1}{2}$). If $M/2<c<M$, $\alpha =a_{0}+a_{1}$, $h_{k}=(M/2)^{k}a_{1}/\alpha $ and the bound from Eq.(\ref{eq:A2}) is valid for $t(M)<t<t(M/2)$, where $t(c)=2\hbar b_{1}/b_{2}$, which leads to $t(M/2)=4\hbar /M$ and $t(M)=(\hbar /M)(2a_{0}+a_{1})/(a_{0}+\frac{1}{4}a_{1})$. For $t<t(M)$, one must use the bound without cut-off. This is illustrated for a particular choice of the $a_{k}$ in Fig.~\ref{fig:3deltas}, which also shows the quartic bounds.

\end{document}